\documentclass[12pt,intlimits,reqno]{amsart}
\usepackage{enumerate}
\usepackage{amssymb}
\usepackage{amsthm}
\usepackage{mathrsfs}
\usepackage{verbatim}
\usepackage{bbm}
\input{xy}
\xyoption{all}
\newtheorem{thm}{Theorem}[section]
\newtheorem{cor}[thm]{Corollary}
\newtheorem{lem}[thm]{Lemma}
\newtheorem{prop}[thm]{Proposition}
\theoremstyle{definition}
\newtheorem{example}[thm]{Example}

\theoremstyle{remark}

\numberwithin{equation}{section}

\newcommand{\norm}[1]{\left\Vert#1\right\Vert}
\renewcommand{\sc}[2]{\langle #1|#2 \rangle}

\newcommand{\abs}[1]{\left\vert#1\right\vert}
\newcommand{\set}[1]{\left\{#1\right\}}
\newcommand{\ket}[1]{\left| #1 \right\rangle}
\newcommand{\R}{\mathbb R}

\newcommand{\No}{\mathbb{N}\cup\{0\}}

\newcommand{\Z}{\mathbb Z}
\newcommand{\C}{\mathbb C}
\newcommand{\N}{\mathbb N}
\renewcommand{\1}{\mathbbm 1}

\renewcommand{\H}{\mathcal{H}}
\renewcommand{\b}{\mathfrak{b}}
\newcommand{\g}{\mathfrak{g}}
\newcommand{\pb}{\{\cdot,\cdot\}}

\renewcommand{\to}{\rightarrow}
\newcommand{\tto}{\longrightarrow}
\newcommand{\prf}[1]{\begin{proof}#1\end{proof}}
\renewcommand{\leq}{\leqslant}
\renewcommand{\geq}{\geqslant}
\renewcommand{\phi}{\varphi}
\renewcommand{\epsilon}{\varepsilon}

\DeclareMathOperator{\Tr}{Tr}
\DeclareMathOperator{\id}{id}

\DeclareMathOperator{\ind}{ind}
\DeclareMathOperator{\ad}{ad}

\DeclareMathOperator{\const}{const}
\DeclareMathOperator{\Ad}{Ad}

\DeclareMathOperator{\Aut}{Aut}

\DeclareMathOperator{\spec}{spec}

\newcommand{\be}{\begin{equation}}
\newcommand{\ee}{\end{equation}}
\newcommand{\bse}{\begin{subequations}}
\newcommand{\ese}{\end{subequations}}
\newcommand{\ben}{\begin{enumerate}}
\newcommand{\een}{\end{enumerate}}
\newcommand{\bit}{\begin{itemize}}
\newcommand{\eit}{\end{itemize}}
\newcommand{\bex}{\begin{example}}
\newcommand{\eex}{\begin{flushright}$\diamondsuit$\end{flushright}\end{example}}

\DeclareMathOperator{\Gr}{Gr_{res}}
\DeclareMathOperator{\Ur}{UL_{res}}
\DeclareMathOperator{\Trr}{Tr_{res}}
\newcommand{\ur}{\mathcal U\mathrm L_\textrm{res}}
\newcommand{\urt}{i\R\oplus\ur}
\newcommand{\urtp}{i\R\oplus\urp}
\newcommand{\urp}{\mathcal U\mathrm L_\textrm{res}^1}

\DeclareMathOperator{\Urot}{\widetilde{UL}_{\textrm{res},0}}

\renewcommand{\u}{\mathcal UL^\infty}
\newcommand{\U}{UL^\infty}

\DeclareMathOperator{\glr}{L_\textrm{res}}
\newcommand{\glrt}{\C\oplus L_\textrm{res}}
\newcommand{\glrtp}{\C\oplus L^1_\textrm{res}}
\newcommand{\glrp}{L^1_\textrm{res}}

\DeclareMathOperator{\GLr}{GL_{res}}

\DeclareMathOperator{\GLro}{GL_{res,0}}
\DeclareMathOperator{\GLrot}{\widetilde{GL}_{res,0}}
\DeclareMathOperator{\GLrk}{GL_{res,k}}
\DeclareMathOperator{\Urk}{UL_{res,k}}
\DeclareMathOperator{\Uro}{UL_{res,0}}

\renewcommand{\Re}{\mathrm{Re}}
\renewcommand{\Im}{\mathrm{Im}}

\newcommand{\ext}{\mathcal E}

\newcommand{\n}{\mathfrak{n}}
\newcommand{\h}{\mathfrak{h}}

\newcommand{\emphh}[1]{{\bf #1}}
\begin{document}

\bigskip

\title[Hierarchy of Hamilton equations on Banach Lie--Poisson spaces\ldots]{Hierarchy of Hamilton equations on Banach Lie--Poisson spaces related to restricted Grassmannian}
\author[Tomasz Goli\'nski, Anatol Odzijewicz]{Tomasz Goli\'nski, Anatol Odzijewicz}
\dedicatory{University in Bia{\l}ystok\\
Institute of Mathematics\\
Lipowa 41, 15-424 Bia{\l}ystok, Poland\\
email: tomaszg@alpha.uwb.edu.pl, aodzijew@uwb.edu.pl
}

\begin{abstract}
Using the Magri method one defines an involutive family of Hamiltonians on Banach Lie--Poisson spaces $\glrtp$ and $\urtp$ which contains the restricted Grassmannian $\Gr$ as a symplectic leaf. The hierarchy of Hamilton equations given by these Hamiltonians is investigated. The operator equations of Ricatti-type are included in this hierarchy. For a few particular cases one gives the explicit solutions.
\end{abstract}

\maketitle

\tableofcontents

\section{Introduction}

In order to formulate geometrically and analytically rigorous language for the theory of infinite dimensional Hamiltonian systems one investigates in \cite{OR} the foundation of Banach Poisson differential geometry. A special place in this theory is occupied by the Banach Lie--Poisson space. Recall that by definition $\b$ is a Banach Lie--Poisson space if its dual $\b^*$ is a Banach Lie algebra such that $\ad_x^*\b\subset\b\subset\b^{**}$ for $x\in\b^*$, where $\ad_x^*:\b^{**}\to\b^{**}$ is dual to the adjoint representation $\ad_x:=[x,\,\cdot\,]:\b^*\to\b^*$. 

Many infinite dimensional physical systems can be considered as a systems on some Banach Lie--Poisson space $\b$ in the Hamilton way 
\be\frac{d}{dt}\rho=-\ad^*_{Dh(\rho)}\rho,\ee
where $\rho\in\b$ and $h\in C^\infty(\b)$, e.g. see \cite{Oind}.

The first aim of this paper is to investigate those Banach Lie--Poisson spaces which are related to the restricted Grassmannian $\Gr$, see \cite{Oext,Ratiu-grass}. The restricted Grassmannian $\Gr$ has its own long story as one of the most important infinite dimensional K\"ahler manifolds in the mathematical physics. It is a set of Hilbert subspaces $W\subset \H$ of a polarized Hilbert space $\H=\H_+\oplus \H_-$ such that the projectors
$P_+:W\to\H_+$ and $P_-:W\to\H_-$ are Fredholm and Hilbert-Schmidt operators respectively.
The geometry of $\Gr$ and its symmetry group play an important role in the quantum field theory \cite{powers,shale-stinespring,wurzbacher,spera-wurzbacher1,thaller}, the loop group theory \cite{segal,sergeev}, and the integration of the KdV and KP hierarchies \cite{sato-sato,segal-wilson,miwa-jimbo-date}.

The second aim is to define and investigate a hierarchy of the Hamilton equations on the Banach Lie--Poisson space $\urtp$ and on its complexification $\glrtp$. This hierarchy is obtained  from the involutive system of Hamiltonians  constructed on $\urtp$ and $\glrtp$ by the Magri method \cite{magri}.

The pair of coupled operator Ricatti equations, see \eqref{ricatti}, belongs to this hierarchy. As one shows in Example \ref{ex:4} the finite dimensional version of the hierarchy provides the non-trivial example of integrable Hamiltonian system.

In our considerations we use the functional analytical methods as well as Banach differential geometric methods. All obtained results are valid in finite dimension case too. 

Since the hierarchy consists of Hamilton equations, the flows preserve symplectic leaves of $\urtp$ and $\glrtp$. In particular case, they preserve $\Gr$, which is one of the symplectic leaves of $\urtp$, see \cite{Ratiu-grass}. We have shown in Example \ref{ex:gr} that after restriction to $\Gr$ the flows linearize in natural complex coordinates.

The central place of the paper is occupied by Section \ref{sec:hier} where (using the Magri method) we construct the infinite hierarchy under consideration. We also discuss its various realizations, see \eqref{h-P+}, \eqref{h-mu}, \eqref{eq-H}, \eqref{eq-x} and \eqref{eq-y}.

In Section \ref{sec:blp-grass} we prepare the material necessary for application of Magri method, i.e. we find explicitly formulas for the coadjoint representation of central extension $\GLrot$ of $\GLro$, the Poisson bracket and Casimirs of the Banach Lie--Poisson spaces $\glrtp$ and $\urtp$.

Finally Section \ref{sec:ex} gives explicit formulas for solutions in some particular cases.

We also include in the paper two Appendices, where we present the Magri method and the theory of extensions of Banach Lie--Poisson spaces.

\section{Banach Lie--Poisson spaces related to restricted Grassmannian}\label{sec:blp-grass}

We investigate the extensions of Banach Lie groups, Banach Lie algebras and Banach Lie--Poisson spaces  
related to restricted Grassmannian $\Gr$. One of the aims of this section is to obtain explicit formulas for adjoint and coadjoint actions of constructed Banach Lie groups and Banach Lie algebras. To this end we apply the methods described in Appendix \ref{ap-a}.

Before that let us recall the definitions of objects we are going to use and fix the notation. For more information see \cite{segal,wurzbacher,sergeev}.

\subsection{Preliminary definitions and notation}

One considers a complex separable Hilbert space with a fixed decomposition onto two orthogonal Hilbert subspaces
\be\label{polarization}\H=\H_+\oplus \H_-.\ee
Let $P_+$ and $P_-$  denote the orthogonal projectors onto $\H_+$ and $\H_-$ respectively. We assume that in general both Hilbert subspaces are infinite dimensional. However we also admit the case when one (or both) of them is finite dimensional. In this case many analytical problems simplify considerably.

In the following we omit the symbols $\H$ and $\H_\pm$ in the notations for various operator algebras and groups and put the subscript $\pm$ if we mean that the operators act in $\H_\pm$. In this way, for example instead of $L^2(\H)$ or  $L^2(\H_+)$ we write $L^2$ or $L^2_+$.

In order to simplify our notation we use the block decomposition 
\begin{align} \label{blocks}P_+AP_+=\left(%
\begin{array}{cc}
  A_{++} & 0\\
  0 & 0\\
\end{array}%
\right),\quad 
P_+AP_-=\left(%
\begin{array}{cc}
  0 & A_{+-}\\
  0 & 0\\
\end{array}%
\right),  \\
P_-AP_+=\left(%
\begin{array}{cc}
  0& 0\\
  A_{-+} & 0\\
\end{array}%
\right),\quad
P_-AP_-=\left(%
\begin{array}{cc}
  0 & 0\\
  0 & A_{--}\\
\end{array}%
\right)\nonumber\end{align}
and we identify the operators
$A_{++}:\H_+\to\H_+$, $A_{--}:\H_-\to\H_-$, $A_{-+}:\H_+\to\H_-$ and $A_{+-}:\H_-\to\H_+$
with $P_+ A P_+$, $P_- A P_-$, $P_-AP_+$, $P_+AP_-$ respectively when there is no risk of confusion.



By $L^p$ we denote the Schatten classes of operators acting in $\H$ equipped with the norm $\norm{\,\cdot\,}_p$. The $L^p$ spaces are ideals in associative algebra $L^\infty$ of bounded operators in $\H$. In particular $L^1$ denotes the ideal of trace-class operators and $L^2$ is the ideal of Hilbert--Schmidt operators. By $L^0\subset L^\infty$ one denotes the ideal of compact operators, which is $\norm\cdot_\infty$-norm closure $\overline{L^p}=L^0$ of any $L^p$ ideal, see \cite{harpe,schatten}.

Let $GL^\infty$ be the Banach Lie group of invertible bounded operators in $\H$. By $\U\subset GL^\infty$ we denote the real Banach Lie group of the unitary operators and its Lie algebra is denoted by $\u$. By $GL^1_+$ we denote the group of invertible operators on $\H_+$ which have a determinant (i.e. they differ from identity by a trace-class operator) and by $SL^1_+$ --- its subgroup which consists of operators with determinant equal to $1$. 
See \cite{reed4,gohberg} for the definition and properties of determinant for this case.

The \emphh{unitary restricted group} $\Ur$ is defined as
\be\Ur:=\{ u\in \U \;|\; [u,P_+]\in L^2\}.\ee
It possesses the Banach Lie group structure given by the embedding
\be \Ur\ni u\mapsto (u,u_{-+})\in \U\times L^2_{+-}.\ee
This structure is not compatible with Banach Lie group structure of $\U$.
The Banach Lie algebra of $\Ur$ is
\be\ur:=\{ x\in L^\infty \;|\; x^+=-x,\; [x,P_+]\in L^2\}\ee
with the norm
\be \norm x_{\textrm{res}}:=\norm{x_{++}}_\infty+\norm{x_{--}}_\infty+\norm{x_{-+}}_2+\norm{x_{+-}}_2,\ee
where $x^+$ is operator adjoint to $x$.
Note that the topology of $\ur$ is strictly stronger than the operator topology on $L^\infty$.

The complexifications of $\Ur$ and $\ur$ are 
$\Ur^\C=\GLr$ and $\ur^\C=\glr$
respectively, where
\be\GLr=\{ g\in GL^\infty \;|\; [g,P_+]\in L^2\}\ee
and
\be\glr=\{ x\in L^\infty \;|\; [x,P_+]\in L^2\}.\ee

\pagebreak

By definition the \emphh{restricted Grassmannian} $\Gr$ consists of Hilbert subspaces $W\subset\H$ such that:
\ben[i)]
\item the projection $P_+$ restricted to $W$ is a Fredholm operator;
\item the projection $P_-$ restricted to $W$ is a Hilbert--Schmidt operator;
\een
see e.g. \cite{segal,wurzbacher}.

The restricted Grassmannian is a Hilbert manifold modelled on the Hilbert space $L^2_{+-}$. The groups $\Ur$ and $\GLr$ act on it transitively. In this way the tangent space to the restricted Grassmannian in the point $\H_+$ can be described as follows 
\be T_{\H_+}\Gr\cong \ur/(\u_+\times \u_-).\ee

Both $\Ur$ and $\GLr$ are disconnected and their connected components are 
$\Urk$ and $\GLrk$, where $g\in\Urk$ and $g\in\GLrk$ iff the Fredholm index $\ind g_{++}$ of the upper left block $g_{++}$ of the operator $g$ is equal to $k$, see \cite{CHO'B,segal}. The maximal connected subgroups $\Uro$ and $\GLro$ will be of special interest. In a similar fashion, connected components of the restricted Grassmannian $\Gr$ are the sets $\Gr_{,k}$ consisting of elements of $\Gr$ such that the index of orthogonal projection $P_+$ restricted to that element is equal to $k$. Let us note that $\Uro$ acts transitively on $\Gr_{,0}$.

\subsection{Extensions of $\GLro$}$\;$\label{ext-glr}

The central object in the following construction is the group $\ext$ defined as
\be \ext:=\{(q,A)\in GL^\infty_+\times\GLro\,|\, A_{++}-q\in L^1_+\}\ee
with pairwise multiplication. The topology and Banach manifold structure on $\ext$ is given by the embedding 
\be(q,A)\mapsto (A_{++}-q,A)\in L^1_+\times\GLr,\ee
see \cite{segal,wurzbacher}.

Let us consider the Banach Lie group extensions presented in the following commutative diagram:
\be\label{group-ext}\xymatrix@=35pt{
 & {\set{1}} & {\set{1}} & {\set{1}}\\
{\set{1}}\ar@{^{(}->}[r]& {\C}^{\times}  \ar@{^{(}->}[r]^-{\iota} \ar[u]& {\GLrot}  \ar[r]^-{\pi} \ar[u]& {\GLro} \ar[r] \ar[u]& {\set{1}}\\
{\set{1}}\ar@{^{(}->}[r]&GL^1_+ \ar@{^{(}->}[r]^-{\iota_1} \ar[u]^{\det}& {\ext} \ar[r]^-{\pi_2}\ar[u]^{\delta}& {\GLro}\ar[u]^{\id}\ar[r] & {\set{1}} \\
{\set{1}}\ar@{^{(}->}[r]& SL^1_+ \ar@{^{(}->}[r]^-{\iota_1} \ar@{^{(}->}[u] &  {SL^1_+\times {\set{\1}}} \ar[r]^-{\pi_2} \ar@{^{(}->}[u] & {\set{1}}\ar@{^{(}->}[u]\\
& {\set{1}}\ar@{^{(}->}[u] & {\set{1}}\ar@{^{(}->}[u] }\ee

The map $\iota_1$ is defined by $\iota_1(q):=(q,\1)$ and the map $\pi_2$ is a projection onto the second component of the Cartesian product $GL^\infty_+\times\GLro$. Thus $\iota_1(SL^1_+)$ is a normal subgroup of $\ext$ and the group $\GLrot$ is defined as the quotient group
\be \GLrot:=\ext/\iota_1(SL^1_+).\ee
The maps $\iota$ and $\pi$ in upper row of diagram \eqref{group-ext} are given as quotients of $\iota_1$ and $\pi_2$ respectively. The map $\delta$ is the quotient map $\ext\to \ext/\iota_1(SL^1_+)=\GLrot$.
In this way all rows and columns in diagram \eqref{group-ext} are exact sequences of Banach Lie groups.


Using the approach described in Appendix \ref{ap-a} we define a local section \eqref{section} of the bundle 
$GL^1_+\to\ext\to\GLro$
by
\be \sigma(A):=(A_{++},A)\ee
for $A\in \GLro$ such that $A_{++}$ is invertible.
The Banach Lie group $\ext$ can be locally identified with $GL^1_+\times_{\Phi,\Omega}\GLro$  through the isomorphism $\Psi:GL^1_+\times_{\Phi,\Omega} \GLro \to \ext$ given in the properly chosen neighborhood of identity by
\be\Psi(n,A)=(nA_{++},A).\ee
The maps $\Phi:\GLro\to \Aut GL^1_+$ and $\Omega:\GLro\times\GLro\to GL^1_+$ defined in general case by \eqref{Phi} and \eqref{Omega} in this case can be expressed locally as follows:
\be \label{Phi_r}\Phi(A)(n)= A_{++} \,n\, A_{++}^{-1},\ee
\be \label{Omega_r}\Omega(A_1,A_2)=A_{1 ++}\, A_{2 ++}\, (A_1 A_2)_{++}^{-1}\ee
for $n\in GL^1_+$, $A,A_1,A_2\in \GLro$ such that $A_{++}, A_{1 ++}, A_{2 ++}$ and $(A_1 A_2)_{++}$ are invertible.

The map $\Phi(A)$ descends to trivial automorphism of $\C^\times$. Thus the Banach Lie group $\GLrot$ can be identified with $\C^\times\times_{\id,\tilde\Omega} \GLro$  for
\be \tilde\Omega:=\det\circ\,\Omega\ee
and it is a central extension of $\GLro$.

\subsection{Extensions of $\glr$}$\;$

The Banach Lie algebra counterpart of the diagram \eqref{group-ext}  is the following:
\be\label{alg-ext}\xymatrix@=35pt{
 & {\set{0}} & {\set{0}} & {\set{0}}\\
{\set{0}}\ar@{^{(}->}[r]& {\C}  \ar@{^{(}->}[r] \ar[u]& {\C\oplus\glr}  \ar[r] \ar[u]& {\glr} \ar[r] \ar[u]& {\set{0}}\\
{\set{0}}\ar@{^{(}->}[r]&L^1_+ \ar@{^{(}->}[r]^-{\iota_1} \ar[u]^{\Tr}& {L^1_+\oplus\glr} \ar[r]^-{\pi_2}\ar[u]^{\Tr_1}& {\glr}\ar[u]^{\id}\ar[r] & {\set{0}} \\
{\set{0}}\ar@{^{(}->}[r]& \mathscr SL^1_+ \ar@{^{(}->}[r]^-{\iota_1} \ar@{^{(}->}[u] &  {\mathscr SL^1_+\oplus \set{0}} \ar[r]^-{\pi_2} \ar@{^{(}->}[u] & {\set{0}}\ar@{^{(}->}[u]\\
& {\set{0}}\ar@{^{(}->}[u] & {\set{0}}\ar@{^{(}->}[u] }\ee
where
\be\mathscr SL^1_+:=\{\rho\in L^1_+\;|\; \Tr \rho=0\}\ee
is the Banach Lie algebra of the group $SL^1_+$.
The Banach Lie algebra $( L^1_+\oplus\glr)/(\mathscr SL^1_+\oplus\set{0})$ of the quotient group $\GLrot$ is naturally identified by $D\Psi(\1,\1)$ with $\glrt$. The map $\Tr_1$ is given by taking trace of the first component of $(\rho,X)\in L^1_+\oplus\glr$.
The direct sums in diagram \eqref{alg-ext} are understood as direct sums of Banach spaces.

Similarly as in the group case, all rows and columns in diagram \eqref{alg-ext} are exact sequences of Banach Lie algebras.


Using the formula \eqref{Ad-ext} with the functions \eqref{Phi_r} and \eqref{Omega_r} we obtain a local formula for the adjoint action
\be\label{Ad-glr+l1}\Ad_{(n,A)}(\rho,X)=\big(nA_{++}(\rho+X_{++})A_{++}^{-1}n^{-1}-(AXA^{-1})_{++},AXA^{-1}\big) \ee
for $(n,A)$ in some open set in $GL^1_+\times_{\Phi,\Omega}\GLro$ and $(\rho,X)\in L^1_+\oplus \glr$.
From \eqref{phi_d} and \eqref{omega_d} we get that
\be \label{phi_r}\phi(X):=[X_{++},\,\cdot\;],\ee
\be \label{omega_r}\omega(X,Y):=-X_{+-}Y_{-+}+Y_{+-}X_{-+}.\ee
The bracket \eqref{ext-bracket} for $\phi$ and $\omega$ given by \eqref{phi_r} and \eqref{omega_r} assumes the form
\be\begin{split}
\label{brack-glr+l1}  [(\rho,X),(\rho',Y)]&=\big([\rho,\rho']+[X_{++},\rho']-[Y_{++},\rho]-\\
&-X_{+-}Y_{-+}+Y_{+-}X_{-+},[X,Y]\big)\end{split}\ee
where $(\rho,X),(\rho',Y)\in L^1_+\oplus\glr$. 
This Banach Lie algebra was presented in \cite{Oext} (up to the sign conventions) as an example of extensions of Banach Lie algebras.

The structure of Banach Lie algebra on $\glrt$ is given by the following function $\tilde\phi$ 
\be \tilde\phi(X)\equiv 0\ee
and the cocycle $\tilde\omega$
\be \label{schwinger}\tilde\omega(X,Y)=-s(X,Y)=-\Tr(X_{+-}Y_{-+}-Y_{+-}X_{-+}),\ee
where $s(X,Y)$ is called Schwinger term, see \cite{schwinger,wurzbacher}.
Thus the adjoint representation of the Lie group $\C^\times\times_{\Phi,\Omega}\GLro$ on $\glrt$ is given by
\be\label{Ad-glrt}\Ad_{(\gamma,A)}(\lambda,X)=\big(\lambda+\Tr(P_+-A^{-1}P_+A),AXA^{-1}\big)\ee
for $(\gamma,A)\in \C^\times\times_{\Phi,\Omega}\GLro$, $(\lambda,X)\in\glrt$.
Moreover, the Lie bracket for $(\lambda,X),(\lambda',Y)\in\glrt$ is the following
\be\label{brack-glrt}  [(\lambda,X),(\lambda',Y)]=\big(-s(X,Y),[X,Y]\big).\ee

Let us note that the formula \eqref{Ad-ext} allows one to express $\Ad$ only locally. However the right hand side of 
\eqref{Ad-glrt} defines some global representation of $\C^\times\times_{\tilde\Phi,\tilde\Omega}\GLro$, which coincides with $\Ad$ on an open neighborhood of $(1,\1)$. However every open neighborhood of the unit element in Banach Lie group generates connected component, and $\C^\times\times_{\tilde\Phi,\tilde\Omega}\GLro$ is connected. Thus the formula \eqref{Ad-glrt} is valid for all $(\gamma,A)\in\C^\times\times_{\tilde\Phi,\tilde\Omega}\GLro$.

\subsection{Extensions of complex Banach Lie--Poisson space $\glrp$}$\;$

In order to find a Banach space predual  to the Banach Lie algebra $\glr$, we define the 
Banach space
\be \glrp:=\{ \mu\in \glr \;|\; \mu_{++}\in L^1_+\;,\;\mu_{--}\in L^1_-\}\ee
with the norm 
\be\label{norm-pre}\norm\mu_*:=\norm{\mu_{++}}_1+\norm{\mu_{--}}_1+\norm{\mu_{-+}}_2+\norm{\mu_{+-}}_2.\ee
Moreover we define the \emphh{restricted trace} $\Trr:\glrp\to\C$ by
\be\Trr \mu :=\Tr(\mu_{++}+\mu_{--}),\ee
for $\mu\in\glrp$. The domain of $\Trr$ is larger than $L^1$ since $L^1\subset \glr$. However for trace-class operators the restricted trace $\Trr$ coincides with the standard trace $\Tr$.
The properties of restricted trace are similar to the properties of the standard trace but one needs to replace $L^\infty$ with $\glr$.

\begin{prop}\label{l1-ideal}
The Banach space $\glrp$ is an ideal (in the sense of commutative algebras) in the Banach space $\glr$. Moreover for $\mu\in\glrp$, $\nu\in\glr$ we have
\be \label{trr-cyclic}\Trr(\mu\nu)=\Trr(\nu\mu).\ee
\end{prop}
\prf{
The conclusion that $\glrp$ is an ideal follows from the fact that $L^1$ and $L^2$ are ideals in $L^\infty$ 
and product of two operators from $L^2$ is trace-class.

To prove the formula \eqref{trr-cyclic} we expand its left hand side 
\be \Trr(\mu\nu)=\Tr(\mu_{++}\nu_{++}+\mu_{+-}\nu_{-+}+\mu_{-+}\nu_{+-}+\mu_{--}\nu_{--}).\ee
By assumptions of the proposition, we conclude that each term under the trace is a trace-class operator. 
Since for $A\in L^\infty$ and $B\in L^1$ or for $A,B\in L^2$ one has
\be \Tr(AB)=\Tr(BA),\ee
we conclude that 
\be \Trr(\mu\nu)=\Trr(\nu\mu).\ee

}

As a corollary of this proposition we get that for $g\in\GLr$ and $\mu\in\glrp$, the operator $\mu g^{-1}$ belongs to $\glrp$ and 
\be \Trr(g\mu g^{-1}) = \Trr(\mu).\ee

Using the pairing between $\mu\in\glrp$, $A\in\glr$ given by
\be \label{pair-glrp}\langle \mu,A\rangle:=\Trr(\mu A)=\Tr(\mu_{++}A_{++})+\Tr(\mu_{+-}A_{-+})+\Tr(\mu_{-+}A_{+-})+\Tr(\mu_{--}A_{--}),\ee
we conclude that the Banach space $\glrp$ is predual $(\glrp)^*\cong \glr$ of $\glr$. This duality can be found in \cite{Oext,Ratiu-grass}.

\begin{prop}\label{stw-norma-pre}
Space $\glrp$ with norm $\norm{\,\cdot\,}_*$ is Banach $*$-algebra.
\end{prop}
\prf{The only point yet to be shown is the inequality
\be \label{norm-pre-mult}\norm{\mu\rho}_*\leq \norm{\mu}_*\norm{\rho}_*\ee
for $\mu,\rho\in \glrp$.  In order to prove it we observe that 
\be \begin{split}
 \norm{\mu\rho}_*&=\norm{(\mu\rho)_{++}}_1+\norm{(\mu\rho)_{+-}}_2+\norm{(\mu\rho)_{-+}}_2+\norm{(\mu\rho)_{--}}_1=\\
&=\norm{\mu_{++}\rho_{++}+\mu_{+-}\rho_{-+}}_1+
\norm{\mu_{++}\rho_{+-}+\mu_{+-}\rho_{--}}_2+\\
&+\norm{\mu_{-+}\rho_{++}+\mu_{--}\rho_{-+}}_2+
\norm{\mu_{--}\rho_{--}+\mu_{-+}\rho_{+-}}_1.\end{split}\ee
Next applying the following inequalities
\be \norm{\rho}_2\leq \norm{\rho}_1\ee 
\be \norm{\rho\mu}_1\leq \norm{\rho}_\infty\norm{\mu}_1\leq\norm{\rho}_1\norm{\mu}_1\ee
for $\rho,\mu\in L^1$ and the inequalities
\be \norm{\rho\mu}_1\leq \norm{\rho}_2\norm{\mu}_2\ee
\be \norm{\rho\mu}_2\leq \norm{\rho}_\infty\norm{\mu}_2\leq\norm{\rho}_2\norm{\mu}_2\ee
for $\rho,\mu\in L^2$,
we obtain
\be\begin{split}\label{prf-norma-pre}
 &\norm{\mu\rho}_*\leq
\norm{\mu_{++}}_1\norm{\rho_{++}}_1+\norm{\mu_{+-}}_2\norm{\rho_{-+}}_2+
\norm{\mu_{++}}_1\norm{\rho_{+-}}_2+\norm{\mu_{+-}}_2\norm{\rho_{--}}_2+\\
&+\norm{\mu_{-+}}_2\norm{\rho_{++}}_1+\norm{\mu_{--}}_1\norm{\rho_{-+}}_2+
\norm{\mu_{--}}_1\norm{\rho_{--}}_1+\norm{\mu_{-+}}_2\norm{\rho_{+-}}_2\leq\\
&\leq \norm\mu_*\norm\rho_*\end{split}\ee

The last inequality in \eqref{prf-norma-pre} follows directly from \eqref{norm-pre}.

}

The Banach space predual to extended Banach Lie algebra $L^1_+\oplus \glr$ is $L^0_+\oplus \glrp$ with natural component-wise pairing
\be \langle (A,\mu),(\rho,X)\rangle=\Tr(A\rho)+\Trr(\mu X).\ee
It follows from the fact that Banach space dual to the ideal of compact operators $L^0_+$ is $L^1_+$, see \cite{takesaki1}.
Analogously, the predual of $\glrt$ is $\C\oplus\glrp$ with the pairing given by
\be \langle (\gamma,\mu),(\lambda,X)\rangle=\gamma\lambda+\Trr(\mu X),\ee
for $\mu\in\glrp$, $X\in\glr$, $\gamma,\lambda\in\C$.

The map $\Tr^*:\C\to L^\infty_+$ dual to $\Tr:L^1_+\to\C$ is given by $\Tr^*(\lambda)=\lambda\1$. Since the ideal of compact operators $L^0_+$ is Banach space predual to $L^1_+$ and $\Tr^*$ does not take values in $L^0_+$, we conclude that $\Tr^*$ cannot be restricted to predual spaces. 
Therefore only horizontal exact sequences in the diagram \eqref{alg-ext} have their predual counterparts
\be\label{ext-glrp}\xymatrix@=35pt{
{\set{0}}\ar@{^{(}->}[r]& {\glrp}  \ar@{^{(}->}[r]^-{\pi_2^*} & {\C\oplus\glrp}  \ar[r]^-{\iota_1^*} & {\C} \ar[r] & {\set{0}}}\ee
\be\xymatrix@=35pt{
{\set{0}}\ar@{^{(}->}[r]& {\glrp} \ar@{^{(}->}[r]^-{\pi_2^*} & {L^0_+\oplus\glrp} \ar[r]^-{\iota_1^*}& {L^0_+}\ar[r] & {\set{0}} },\ee
where the map $\pi_2^*$ is an injection into the second argument and the map $\iota_1^*$ is the projection onto the first component of the respective direct sums.

It follows from \eqref{coad-ext} and from \eqref{Phi_r}, \eqref{Omega_r} that the coadjoint representation of Banach Lie group
$GL^1_+\times_{\Phi,\Omega}\GLro$ 
on predual Banach space $L^0_+\oplus\glrp$ is given by
\be \label{coAd-glr+l1}\Ad^*_{(n,A)}(\tau,\mu)=(A_{++}^{-1}n^{-1}\tau nA_{++},
A^{-1}_{++}n^{-1}\tau n A_{++}-A^{-1}\tau A+A^{-1}\mu A)\ee
for $(n,A)\in GL^1_+\times_{\Phi,\Omega}\GLro$ and $(\tau,\mu)\in L^0_+\oplus\glrp$.
Similarly, the coadjoint representation of $\C^\times\times_{\id,\tilde\Omega}\GLro$ on $\C\oplus\glrp$ is the following
\be \label{coAd-glrt}\Ad^*_{(\lambda,A)}(\gamma,\mu)=(\gamma,A^{-1}\mu A+\gamma (P_+-A^{-1}P_+A))\ee
where $(\lambda,A)\in \C^\times\times_{\id,\tilde\Omega}\GLro$ and $(\gamma,\mu)\in\C\oplus\glrp$. 

Let us also note that these coadjoint representations preserve 
the Banach subspaces $L^0_+\oplus\glrp\subset (L^1_+\oplus\glr)^*$ and 
$\C\oplus\glrp\subset \C\oplus\glr^*$ respectively
\be \Ad^*_{(n,A)}(L^0_+\oplus\glrp)\subset L^0_+\oplus\glrp\ee
\be \Ad^*_{(\lambda,A)}(\C\oplus\glrp)\subset \C\oplus\glrp.\ee

The coadjoint representation of the Banach Lie algebra $L^1_+\oplus\glr$ on its predual $L^0_+\oplus\glrp$ is the following
\be \ad^*_{(\rho,X)}(\tau,\mu)=([-\rho,\tau]-[X_{++},\tau],-[X,\mu]-[\rho,\tau]-\tau X_{+-} + X_{-+}\tau)\ee
where $(\rho,X)\in L^1_+\oplus\glr$, $(\tau,\mu)\in L^0_+\oplus\glrp$. 

The coadjoint representation of the Banach
Lie algebra $\glrt$ on $\C\oplus\glrp$ is given by
\be \label{coad-glrtp}\ad^*_{(\lambda,X)}(\gamma,\mu)=(0,-[X,\mu]-\gamma(X_{+-} - X_{-+}))\ee
where $(\lambda,X)\in\glrt$, $(\gamma,\mu)\in \C\oplus\glrp$.

We observe that the conditions \eqref{coad-pres} are satisfied for both extensions, thus the Banach spaces 
$L^0_+\oplus\glrp$ and $\C\oplus\glrp$ are Banach Lie--Poisson spaces.
The Poisson bracket for $F,G\in C^\infty(\glrtp)$ is obtained from the general formula \eqref{ext-pb} and  is given by
\be \label{glrtp-pb}\{F,G\}(\gamma,\mu)=\langle \mu,[D_2 F(\gamma,\mu),D_2 G(\gamma,\mu)]\rangle-\gamma s(D_2 F(\gamma,\mu),D_2 G(\gamma,\mu)), \ee
where $D_2$ denotes partial Fr\'echet derivative with respect to the second variable.

\subsection{Extensions of real Banach Lie--Poisson space $\urp$}$\;$

In the previous  subsections we were considering the extensions of the complex linear restricted group $\GLro$, its Banach Lie algebra $\glr$ and the complex Banach Lie--Poisson space $\glrp$, which is Banach predual of $\glr$. As we mentioned above they are complexifications of $\Uro$, $\ur$ and 
\be (\ur)_*\cong\urp:=\{\mu\in\glrp \;|\; \mu^+=-\mu\}\ee
respectively. The pairing between elements of $\ur$ and $\urp$ as in the complex case is given by \eqref{pair-glrp}.

By similar construction as for $\GLro$, we obtain the central extension of $\Uro$ by $U(1)$
\be\label{Uext}\xymatrix@=35pt{
{\set{1}}\ar@{^{(}->}[r]& {U(1)} \ar@{^{(}->}[r]& {\Urot} \ar[r] & {\Uro}\ar[r] & {\set{1}} }.\ee
Namely we define the real Banach Lie group
\be U\ext:=\{(A,q)\in \ext \,|\, A\in \Uro,\; q\in \U_+\},\ee
which complexification $U\ext^\C$ is $\ext$. The group $\Urot$ in \eqref{Uext} is defined as
\be \Urot:=U\ext/\iota_1(SUL^1_+)),\ee
where $SUL^1_+:=SL^1_+\cap \U$ and $\iota_1(q):=(q,\1)$.

Restricting the Schwinger term \eqref{schwinger} to $\ur$ we obtain the central extension
\be\label{uext}\xymatrix@=35pt{
{\set{0}}\ar@{^{(}->}[r]& {i\R} \ar@{^{(}->}[r]& {\urt} \ar[r] & {\ur}\ar[r] & {\set{0}} }\ee
of the real Banach Lie algebra $\ur$. The exact sequence of Banach Lie--Poisson spaces predual to \eqref{uext} is
\be\label{upext}\xymatrix@=35pt{
{\set{0}}\ar@{^{(}->}[r]& {i\R} \ar@{^{(}->}[r]& {\urtp} \ar[r] & {\urp}\ar[r] & {\set{1}} }.\ee
The complexification of \eqref{upext} gives \eqref{ext-glrp}. All expressions obtained above, including the ones for the Poisson bracket \eqref{glrtp-pb} and coadjoint representation \eqref{coAd-glrt}, \eqref{coad-glrtp} are valid for the real case if one assumes that $\bar\gamma=-\gamma$ and $\mu^+=-\mu$.

\section{Hierarchy of Hamilton equations on Banach Lie--Poisson spaces $\glrtp$ and $\urtp$}
\label{sec:hier}
In this section we use the Magri method, see \cite{magri}) to introduce the hierarchy of the Hamiltonian systems on the Banach Lie--Poisson spaces $\glrtp$ and $\urtp$, which were investigated in Section \ref{sec:blp-grass}. Short description of Magri method is presented in Appendix \ref{ap:magri}.

To this end we define for any $k\in\N$ the function
\be \label{cas}I^k(\gamma,\mu):=\Trr\big( (\mu-\gamma P_+)^{k+1}-(-\gamma)^{k}(\mu-\gamma P_+)\big)\ee
on $\glrtp$. Note that the expression under $\Trr$ is a polynomial in variable $\mu$ without a free term, thus from Proposition \ref{l1-ideal} it follows that the function $I^k$ is well defined.

We observe that $I^k$ is invariant with respect to the coadjoint representation \eqref{coAd-glrt}
\be\begin{split}
 &I^k(\Ad^*_{[n,A]} (\gamma,\mu))=I^k(\gamma,A^{-1}\mu A+\gamma(P_+-A^{-1}P_+A))=\\
&=\Trr\big( A^{-1}(\mu-\gamma P_+ + \gamma A P_+ A^{-1}-\gamma A P_+ A^{-1})^{k+1} A -\\
&-A^{-1}(-\gamma)^{k}(\mu-\gamma P_+ + \gamma A P_+ A^{-1}-\gamma A P_+ A^{-1})A\big)=\\
&=\Trr\big((\mu-\gamma P_+)^{k+1} -(-\gamma)^{k}(\mu-\gamma P_+ )\big)=I^k(\gamma,\mu).
\end{split}\ee
Thus the functions $I^k$, $k\in\N$,  are Casimirs
\be \{I^k,\,\cdot\,\}=0\ee
for Poisson bracket \eqref{glrtp-pb}. 
Note that the coordinate function $\gamma$ is a Casimir too.


Observing that Poisson brackets for $F,G\in C^\infty(\glrtp)$ given by
\be \{F,G\}_1(\gamma,\mu):=\langle \mu,[D_2 F(\gamma,\mu),D_2 G(\gamma,\mu)]\rangle\ee
 and by
\be \{F,G\}_2(\gamma,\mu):=-\gamma\, s(D_2 F(\gamma,\mu),D_2 G(\gamma,\mu))\ee
are compatible, we introduce a Poisson pencil 
\be\begin{split} \label{glrtp-pen}
&\{F,G\}_\epsilon(\gamma,\mu):=\{F,G\}_1(\gamma,\mu)+\epsilon\{F,G\}_2(\gamma,\mu)= \\
&\langle \mu,[D_2 F(\gamma,\mu),D_2 G(\gamma,\mu)]\rangle-\epsilon \gamma \, s(D_2 F(\gamma,\mu),D_2 G(\gamma,\mu))\end{split}\ee
on $\glrtp$. Compatibility of $\pb_1$ and $\pb_2$ follows from the fact that the Poisson tensor for $\pb_2$ is constant with respect to the variable $\mu$ and $\pb_1$ depends only on the derivations with respect to the variable $\mu$.

Due to the last equality in \eqref{glrtp-pen} the Casimirs for $\pb_\epsilon$ are:
\be \label{casimirs} I^k_\epsilon(\gamma,\mu)=\Trr\big( (\mu-\epsilon\gamma P_+)^{k+1}-(-\epsilon\gamma)^{k}(\mu-\epsilon\gamma P_+)\big)\ee
where $k\in\N$. According to Magri method, we expand these Casimirs with respect to the parameter $-\epsilon\gamma$
\be \label{W-expansion}I^k_\epsilon(\gamma,\mu)=\sum_{n=0}^{k-1}(-\epsilon\gamma)^n\Trr W_n^{k+1}(\mu)+(-\epsilon\gamma)^{k}\Trr\big(W^{k+1}_{k}(\mu)-\mu\big),\ee
where the operators $W_n^k(\mu)$ are polynomials in operator arguments $\mu$ and $P_+$ defined by the equality
\be \label{szereg}(\mu+\lambda P_+)^k=\sum_{n=0}^k\lambda^n W_n^k(\mu),\qquad \lambda\in\R. \ee
In this way from \eqref{magri_inv} it follows that we obtain a family
\bse\label{calki}\begin{align}
&h_n^k(\gamma,\mu)=\gamma^n \Trr W_n^{k+1}(\mu),\quad 0\leq n\leq k-1,\\
&h_{k}^k(\gamma,\mu)=\gamma^{k}\Trr\big(W_{k}^{k+1}-\mu\big)
 \end{align}\ese
of Hamiltonians in involution
\be \{h^k_n,h^l_m\}_\epsilon=0\ee
with respect to the brackets $\pb_\epsilon$ for $\epsilon\in\R$. In the particular case $\epsilon=1$ they are in involution with respect to the bracket $\pb$ given by \eqref{glrtp-pb}.


Let us now investigate the infinite system of the Hamilton equations on the Banach Lie--Poisson space $\glrtp$
\be \label{ham-ad}\frac\partial{\partial t_n^k} (\gamma,\mu)=-\ad^*_{(D_1 h_n^k(\gamma,\mu),D_2 h_n^k(\gamma,\mu))}(\gamma,\mu)\ee
defined by the hierarchy of Hamiltonians $h_n^k$, $k\in\N$, $n=0,1,\ldots,k$.
Using the explicit form of coadjoint action \eqref{coad-glrtp} we observe that the equations \eqref{ham-ad} assume the form
\bse\label{h}\begin{align}\frac\partial{\partial t_n^k} \gamma=&0\\
\frac\partial{\partial t_n^k} \mu=&-[\mu,D_2 h_n^k(\gamma,\mu)]+
\gamma(P_+D_2 h_n^k(\gamma,\mu)P_--P_-D_2 h_n^k(\gamma,\mu)P_+).
\end{align}\ese

In \eqref{h} the real parameter $t^k_n$ parametrizes the Hamiltonian flow generated by $h_n^k$. In order to compute $D_2h_n^k(\gamma,\mu)$ we apply the partial derivative operator $D_2$ to both sides of the equality \eqref{W-expansion}. 
Since 
\be D_2 I^k_\epsilon(\gamma,\mu)= (k+1)(\mu-\epsilon\gamma P_+)^k-(-\epsilon\gamma)^k\1\ee
we get that
\bse \label{dh}\be D_2 h_n^k(\gamma,\mu)=(k+1)\gamma^n W^k_n(\mu)\ee
for $0\leq n \leq k-1$, and
\be D_2 h_k^k(\gamma,\mu)=\gamma^k \left((k+1)W^k_k(\mu)-\1\right)
.\ee\ese

Substituting \eqref{dh} into \eqref{h} we obtain
\be\label{h-W} \frac\partial{\partial t_n^k}\mu 
= -(k+1)\gamma^n [\mu-
\gamma P_+, W_n^k(\mu)]\ee
for $0\leq n \leq k$.

From \eqref{szereg} we obtain the following recurrence rules
\be \label{W-recc}\begin{aligned}W_n^{k+1}(\mu)=W^k_n(\mu)\mu+W_{n-1}^k(\mu)P_+\\
		W_n^{k+1}(\mu)=\mu W^k_n(\mu)+P_+W_{n-1}^k(\mu)\end{aligned}\ee
which yield the commutation relation
\be \label{W-comm}[\mu,W_n^k(\mu)]+[P_+,W_{n-1}^k(\mu)]=0,\ee
where $0\leq n\leq k$ and we put $W^k_{-1}:=0$. Using \eqref{W-comm} we express the Hamilton equations \eqref{h-W} in the following two ways
\be\label{h-mu} \frac\partial{\partial t_n^k}\mu = -(k+1)\gamma^n [\mu, W_n^k(\mu)+\gamma W_{n+1}^k(\mu)]\ee
\be \label{h-P+}\frac\partial{\partial t_n^k}\mu = (k+1)\gamma^n [P_+, 
\gamma W_n^k(\mu)+W^k_{n-1}(\mu)],\ee
where $0\leq n\leq k$. Rewriting \eqref{h-P+} in the block form \eqref{blocks} we obtain
\be \label{diag-block}\frac\partial{\partial t_n^k}\,\mu_{++}=0 \qquad \frac\partial{\partial t_n^k}\,\mu_{--}=0\ee
and
\be\label{offdiag-block}\left\{\begin{aligned}
&\frac\partial{\partial t_n^k}\,\mu_{+-}=(k+1)\gamma^n P_+\big(\gamma W^k_n(\mu) + W^k_{n-1}(\mu)\big)P_-\\
&\frac\partial{\partial t_n^k}\,\mu_{-+}=-(k+1)\gamma^n P_-\big(\gamma W^k_n(\mu) + W^k_{n-1}(\mu)\big)P_+
\end{aligned}\right.\ee
Let us observe that the equation \eqref{h-mu} is a Hamilton equation for the Poisson bracket $\pb_1$ and the Hamiltonian $h^k_n+h^k_{n+1}$, while the equation \eqref{h-P+} is a Hamilton equation for the Poisson bracket $\pb_2$ and the Hamiltonian $h^k_n+h^k_{n-1}$. From \eqref{diag-block} we conclude that diagonal blocks $\mu_{++}$ and $\mu_{--}$ are invariants for all Hamiltonian flows under consideration. 

The symplectic leaves for $\glrtp$ and $\urtp$ with Poisson bracket $\pb_2$ are the affine spaces obtained by shifting the vector spaces $L^2_{+-}\oplus L^2_{-+}$ or $L^2_{+-}$ respectively by diagonal blocks $\mu_{++}$ and $\mu_{--}$. This fact explain why we have obtained additional integrals of motion \eqref{diag-block}.

The equations \eqref{h-W} and \eqref{h-mu} are in Lax form. 
In the first case it is an equation on $\mu-\gamma P_+$, while it is on $\mu$ in the other.

Let us calculate several operators $W^k_n(\mu)$. We can do this by iterating the recurrence \eqref{W-recc}. The result read as:
\begin{align}
\label{W-l0}W^k_k&=P_+\\
\label{W-l1}W^k_{k-1}&=\mu P_++P_+\mu + (k-2)P_+\mu P_+\qquad \qquad k\geq 2\\
W^k_{k-2}&=\mu^2 P_++\mu P_+ \mu + P_+ \mu^2 + (k-3)\left(P_+\mu^2 P_+ +P_+\mu P_+\mu+\mu P_+\mu P_+\right)+\nonumber\\
&\label{W-l2}+\frac{(k-3)(k-4)}2 P_+ \mu P_+ \mu P_+ \qquad \qquad k\geq 4\\
\vdots\nonumber\\
W^k_1&=P_+\mu^{k-1}+\mu P_+ \mu^{k-2}+\ldots+\mu^{k-1}P_+\\
 W^k_0&=\mu^k\\
\end{align}
It is obvious that the Hamiltonians $h_n^k$ are functionally interdependent and it implies the interdependence of $t_n^k$--flows given by \eqref{offdiag-block}. The above formulas suggest to introduce the homogeneous polynomials
\be H_n^l(\mu):=\sum^1_{i_0,i_1,\ldots i_l=0 \atop i_0+\ldots +i_l=n} P_+^{i_0}\mu P_+^{i_1}\mu\ldots \mu P_+^{i_l}\ee
of the degree $l\in\N$ in the  operator variable $\mu\in \glrp$, where $n\leq l+1$. These polynomials are linearly independent and 
they satisfy the recurrences
\bse\label{H-recc}\be H^{l+1}_{n+1}(\mu)=P_+\mu H_n^l(\mu)+\mu H^l_{n+1}(\mu)\ee
for $n\leq l$, $l\in \N$ and
\be H^{l+1}_{l+2}(\mu)=P_+\mu H_{l+1}^l(\mu)\ee\ese
for $l\in\N$.

\begin{prop}\label{prop1}
Polynomials $W^k_n$ are linear combinations of the homogeneous polynomials $H^l_n$ 
\be \label{W-H}W^k_{k-l}(\mu)=\sum_{n=1}^{l+1} \max\{0,p_n^l(k)\} H_n^l(\mu)\ee
for $l<k$  and
\be W^k_0(\mu)=H_0^k(\mu),\ee
where $p_n^l\in \R_{n-1}[x]$ are polynomials of degree $n-1$ that are defined by the recurrences:
\be \label{p-rec1}p^l_{n+1}(k)=\sum_{i=l+1}^{k-1} \max\{0,p_n^l(i)\},\ee
\be \label{p-rec2}p^{l+1}_n(k)=p^l_n(k-1)\ee
with initial condition $p^l_1(k)=1$.
\end{prop}
\prf{We prove the formula \eqref{W-H} by induction with respect to $l$. From recurrence \eqref{p-rec1} we infer  that $p^1_2(k)=k-2$ and thus from \eqref{W-l0} and \eqref{W-l1} we see that formula \eqref{W-H} is satisfied for $l=0$ and $l=1$.

From \eqref{W-recc} we conclude that
\be \label{W-re}W^k_{k-l}(\mu)=\mu W^{k-1}_{k-l}(\mu)+P_+\mu\sum_{i=1}^{k-l} W^{k-i-1}_{k-l-i}(\mu).\ee
We apply \eqref{W-re} to $W^k_{k-l}$ assuming that \eqref{W-H} is satisfied for $l-1$ and obtain
\be\begin{split}
 &W^k_{k-l}(\mu)=\mu \sum_{n=1}^l \max\{0,p_n^{l-1}(k-1)\} H^{l-1}_n(\mu)+\\
&+P_+\mu\sum_{i=1}^{k-l-1}\sum_{n=1}^l\max\{0,p_n^{l-1}(k-i-1)\}H^{l-1}(\mu)+P_+\mu H^{l-1}_0(\mu)\end{split}\ee

Changing the order of summation and using recurrences \eqref{p-rec1} and \eqref{p-rec2} we get
\be\begin{split}
 &W^k_{k-l}(\mu)=\mu \sum_{n=0}^{l-1} \max\{0,p_{n+1}^{l-1}(k-1)\} H^{l-1}_{n+1}(\mu)+\\
&+P_+\mu \sum_{n=1}^l\max\{0,p_{n+1}^{l-1}(k-1)\}H^{l-1}_n(\mu)+P_+\mu H^{l-1}_0(\mu).
\end{split}\ee
By rearranging terms in the sums and applying \eqref{H-recc} we end up with \eqref{W-H}.

Direct check shows that relations \eqref{p-rec1} and \eqref{p-rec2} are compatible. The fact that $\deg p_n^l=n-1$ follows from \eqref{p-rec1}.

} 

From Proposition \ref{prop1} we conclude:
\begin{cor}$\;$

\ben[i)]\item The dimension of the complex vector space spanned by $\{W^k_{k-l}\}_{k=l+1}^\infty$ is equal to $l+1$ and
$\{H^l_n\}_{n=1}^{l+1}$ is a basis of this space;
\item Using \eqref{W-H} one can express $H^l_n$, $0\leq n\leq l+1$ as a finite linear combination of $W^k_{k-l}$, where $k\geq l+1$.
\een
\end{cor}
\prf{$\;$

\ben[i)]
\item From \eqref{W-H} it follows that all elements of the set $\{W^k_{k-l}\}_{k=l+1}^\infty$ are linear combinations 
of $\{H^l_n\}_{n=1}^{l+1}$. Let us also note that the polynomials $p_n^l$ assume positive values $p_n^l(k)>0$ for $k$ large enough
and that the set $\{p_n^l\}_{n=1}^{l+1}$ spans an $l+1$-dimensional vector space. It concludes the proof.
\item This statement is a consequence of $i)$.
\een
}

Introducing new variables $\tau^l_n\in\R$ through the linear combination
\be \label{tau}t^k_{k-l}=(k-1)\gamma^{k-l}\sum_{n=1}^{l+1} \max\{0,p_n^l(k)\} \tau_n^l,\ee
we rewrite the hierarchy \eqref{h-W} in the equivalent form
\be \label{eq-H}\frac\partial{\partial \tau_n^l}\mu =  [\mu-\gamma P_+, H_n^l(\mu)]\ee
where $l\in\N$ and $n=1,\ldots, l+1$.

Let us make explicit several equations from hierarchy \eqref{eq-H}. 
For $H_0^k$ and $H_1^k$ in block notation \eqref{blocks} we obtain
\be\label{h0}\left\{\begin{aligned}
&\frac\partial{\partial \tau_0^k}\,\mu_{+-}=-\gamma(\mu^{k})_{+-}\\
&\frac\partial{\partial \tau_0^k}\,\mu_{-+}=\gamma (\mu^{k})_{-+}
\end{aligned}\right.\ee
and
\be\label{h1}\left\{\begin{aligned}
&\frac\partial{\partial \tau_1^k}\,\mu_{+-}=-(\mu^{k+1})_{+-}-\gamma\sum_{i=0}^{k-1}(\mu^{i})_{++}(\mu^{k-i})_{+-}\\
&\frac\partial{\partial \tau_1^k}\,\mu_{-+}=(\mu^{k+1})_{-+}+\gamma\sum_{i=1}^{k}(\mu^{i})_{-+}(\mu^{k-i})_{++}
\end{aligned}\right.\ee
respectively. For $k=1$ and $k=2$ we get from \eqref{h0} linear equations and for $k=3$ we obtain from \eqref{h0} a pair of coupled operator Ricatti-type equations
\be\label{ricatti}\left\{\begin{aligned}
&\frac\partial{\partial \tau_0^3}\,\mu_{+-}=-\gamma\big((\mu_{++})^2\mu_{+-}+\mu_{+-}\mu_{-+}\mu_{+-}+\mu_{++}\mu_{+-}\mu_{--}+\mu_{+-}(\mu_{--})^2\big)
\\
&\frac\partial{\partial \tau_0^3}\,\mu_{-+}=\gamma\big(\mu_{-+}(\mu_{++})^2+\mu_{--}\mu_{-+}\mu_{++}+\mu_{-+}\mu_{+-}\mu_{-+}+(\mu_{--})^2\mu_{-+}\big)
\end{aligned}\right..\ee

Let us recall that blocks $\mu_{++}$ and $\mu_{--}$ are constant with respect to all flows. Moreover if we assume that $\mu_{++}=0$ or $\mu_{--}=0$ then the equations \eqref{eq-H} become linear.


After certain modification we can also consider the hierarchy of equations \eqref{eq-H} on the real Banach Lie--Poisson $\urtp$. To this end we have to modify the Hamiltonians $h^k_n$ is such way that they will assume real values when restricted to $\urtp$. In consequence we obtain the following equations
\be \label{eq-re}\frac\partial{\partial \tau_n^l}\mu =   i^{l+1} [\mu-\gamma P_+, H_n^l(\mu)]\ee
where $\mu\in\urp$, $\gamma\in i\R$.


Now we  express the Hamiltonian hierarchy \eqref{h-W} in a more compact and elegant form. To this end let us define the "generating" Hamiltonian for the Hamiltonians \eqref{calki}
\be \label{h_kappa}h_{\kappa,\lambda}(\gamma,\mu):=
\sum_{k=1}^\infty\frac1{k+1}\kappa^k\sum_{n=0}^{k+1}\lambda^n  h^k_n(\gamma,\mu),\ee
where $\kappa,\lambda\in \R$. In order to show that the series of functions \eqref{h_kappa} is convergent on some non-empty open subset of $\glrtp$ we observe that
\be \label{genf}\sum_{n=0}^{k+1}\lambda^n  h^k_n(\gamma,\mu)=\Trr\left((\mu+\gamma\lambda P_+)^{k+1}-(\gamma\lambda)^k(\mu+\gamma\lambda P_+)\right).\ee
The equality \eqref{genf} follows  from \eqref{calki} and \eqref{szereg}. Next, let us prove the following lemma.

\begin{lem}
 One has
\be \norm{(\mu+\beta P_+)^{k+1}-\beta^k(\mu+\beta P_+)}_*\leq (\norm\mu_*+\abs\beta)^{k+1}-\abs{\beta}^k(\norm\mu_*+\abs\beta),\ee
where $\beta\in\C$ and $\mu\in\glrp$.
\end{lem}
\prf{We expand the left hand side and apply triangle inequality and Proposition \ref{stw-norma-pre}. Moreover we note that $\norm{\nu P_+}_*\leq \norm\nu_*$ for $\nu\in\glrp$. In this way we get
\be \norm{(\mu+\beta P_+)^{k+1}-\beta^k(\mu+\beta P_+)}_*\leq \sum_{i=1}^{k+1} \left({k+1 \atop i}\right)\norm\mu_*^i\abs\beta^{k-i+1}-\abs\beta^k\norm\mu_*.\ee
By adding and subtracting the term $\abs\beta^{k+1}$ and collecting terms we obtain the right hand side.

}

From this lemma and Proposition \ref{stw-norma-pre} we conclude:
\begin{prop}
One has
\be h_{\kappa,\lambda}(\gamma,\mu)=\Trr\left(\frac{1}\kappa \log \big(\1-\kappa (\mu+\gamma\lambda P_+)\big)-(\mu+\gamma\lambda P_+)\frac{\log(1-\kappa\lambda\gamma)}{\kappa\lambda\gamma}\right)\ee
and
\be\label{h-gen-inv}\{h_{\kappa,\lambda},h_{\kappa',\lambda'}\}=0\ee
for $\abs\kappa(\norm\mu_*+\abs{\lambda\gamma})<1$ and $\abs{\kappa'}(\norm\mu_*+\abs{\lambda'\gamma})<1$, where the Poisson bracket in \eqref{h-gen-inv} is given by \eqref{glrtp-pb}.
\end{prop}

Now we find the explicit form of the Hamilton equation
\be \frac\partial{\partial t_{\kappa,\lambda}} (\gamma,\mu)=-\ad^*_{D h_{\kappa,\lambda}(\gamma,\mu)}(\gamma,\mu)\ee
generated by the Hamiltonian \eqref{h_kappa}. Using \eqref{coad-glrtp} we obtain
\bse\begin{align}\frac\partial{\partial t_{\kappa,\lambda}} \gamma=&0\\
\frac\partial{\partial t_{\kappa,\lambda}} \mu=&-[\mu,D_2 h_{\kappa,\lambda}(\gamma,\mu)]+
\gamma(P_+D_2 h_{\kappa,\lambda}(\gamma,\mu)P_--P_-D_2 h_{\kappa,\lambda}(\gamma,\mu)P_+).
\end{align}\ese
Since
\be D_2 h_{\kappa,\lambda}(\gamma,\mu)=\big(\1-\kappa(\mu+\lambda\gamma P_+)\big)^{-1}-\frac{\log(1-\kappa\lambda\gamma)}{\kappa\lambda\gamma}\ee
we get
\be \label{eq-x}\frac\partial{\partial t_{\kappa,\lambda}}  x =-\alpha[P_+,(1-x)^{-1}],\ee
where
\be x:=\kappa(\mu+\lambda\gamma P_+),\ee
and $\alpha:=\kappa(1+\lambda)\gamma$. Replacing $x$ in \eqref{eq-x} by $y:=(1-x)^{-1}$ we get the hierarchy of equations
\be \label{eq-y}\frac\partial{\partial t_{\kappa,\lambda}}  y =\alpha[y,yP_+y],\qquad \lambda,\kappa\in \C\ee
equivalent to the hierarchy \eqref{h-W}.

In this paper we don't intend to address the problem of finding general solutions for the considered Hamiltonian systems but in the next section we will present several examples of solutions.

Let us also observe that if $\H_+$ is finite dimensional, then the operator $\mu-\gamma P_+$ has a discrete spectrum. The formula \eqref{coAd-glrt} shows that orbits of coadjoint action of group $\GLrot$ coincide with orbits of standard coadjoint action of $\GLro\subset GL^\infty$ shifted by $\gamma P_+$. Thus 
one can use the spectrum $\spec(\mu-\gamma P_+)$  to distinguish partially symplectic leaves of the Banach Lie--Poisson space $\urtp$, i.e. if $\spec(\mu_1-\gamma P_+)\neq \spec(\mu_2-\gamma P_+)$ then they belong to different symplectic leaves. If the dimension of $\H_+$ is infinite then the shift $\mu-\gamma P_+$ of the operator $\mu\in\glrp$ by the operator $\gamma P_+$ is not an element of $\glrp$, but of $\glr$. However from Weyl's criterion (see \cite{reed1}) we deduce that set $\spec(\mu-\gamma P_+)\setminus\{0,-\gamma\}$ is discrete. In such way, also if $\H_+$ is infinite dimensional, one can use elements of $\spec(\mu-\gamma P_+)$ for the partial indexation of the symplectic leaves. The problem of description of these leaves is complicated, see \cite{Ratiu-grass} for more information.

\section{Examples of solutions}\label{sec:ex}

In this section we present several examples of explicit solutions of equations \eqref{eq-H} in some particular cases.

\begin{example}[restricted Grassmannian]$\;$\label{ex:gr}

The connected component $\Gr_{,0}$ of the restricted Grassmannian $\Gr$ can be identified with the coadjoint orbit $\mathcal O_\gamma$ of the group $\Urot$ in the Banach Lie--Poisson space $\urtp$ generated from point $(\gamma,0)$, see \cite{Ratiu-grass}. Namely from \eqref{coAd-glrt} we see that
\be \Ad^*_{(\lambda,g)}(\gamma,0)=(\gamma,\gamma (P_+-g^{-1}P_+g))\ee
for $(\lambda,g)\in U(1)\times_{\id,\tilde\Omega}\Uro$ 
and $\gamma\in i\R$. It suggests to define a map
\mbox{$\iota_\gamma:\Gr_{,0}\to \mathcal O_\gamma$}
 in the following way
\be \iota_\gamma(W):=(\gamma,\gamma(P_+-P_W)).\ee
Since $\Uro$ acts transitively on $\Gr_{,0}$, we see that $\iota_\gamma$ maps $\Gr_{,0}$ bijectively on $\mathcal O_\gamma$.

Let us introduce homogeneous coordinates on some open subset in $\Gr$. To this end we fix a basis $\{\ket n\}$, $n\in\Z$ in $\H$, such that $\ket n$ for $n<0$ spans $\H_-$ and for $n\geq 0$ spans $\H_+$. Let us fix a basis $w_1,w_2,\ldots$ in a subspace $W\in\Gr$ and put the coefficients of $w_k$ in the basis $\{\ket n\}_{n\in\Z}$ in the matrix form
\be \left({\alpha \atop \beta}\right):=( \sc{n}{w_k} )_{n\in\Z,k\in\N},\ee
where $\alpha,\beta$ are blocks obtained for $n\geq 0$ and $n<0$ respectively. Let us consider a $W\in\Gr$ such that there exists an orthonormal basis $\{w_k\}_{k\in\N}$ such that $\alpha$ is invertible. Then we define
\be z:=\beta\alpha^{-1}.\ee
Definition of $z$ is independent on the choice of basis $\{w_k\}_{k\in\N}$.

The matrix of the projector $P_W$ reads as
\be (\sc{n}{P_W k})_{n,k\in\Z} =\left({\alpha \atop \beta}\right)\left( \alpha^+\;\beta^+\right)\ee
Thus $\iota_\gamma(W)$ assumes the following form
\be \iota_\gamma(W)=\begin{pmatrix}
(\1+z^+z)^{-1}-\1 & (\1+z^+z)^{-1}z^+\\
z(\1+z^+z)^{-1}&z(\1+z^+z)^{-1}z^+     \end{pmatrix},\ee
where we consider $z$ as an operator $z:\H_+\to\H_-$.

Hamilton equations \eqref{h} can be written in terms of $z$. Due to \eqref{diag-block} we note that $z^+z$ is constant. Any term obtained as a product of blocks is either constant or linear in $z$.
\end{example}

\begin{example}[Vector case]$\;$\label{eq:vector}

Let us consider a particular case of the equations given above. We assume that 
$\dim\H_+=1$. We introduce the block notation for elements $\mu\in \glrp$
\be \mu=\begin{pmatrix}
a & v^+\\
w&A
        \end{pmatrix},\ee
where $a\in\C$, $A\in L^\infty(\H_-)$, $v,w\in L^2(\C,\H_-)\cong \ell^2$. We consider equations \eqref{eq-H} as non-linear equations for two vectors $v,w$ coupled by interaction depending on constants $a$ and $A$.
Non-linear behavior is due to the terms of the type $\sc v{A^l w}$.

First of all let us remark that in general case all functions $h_l^k$ are linear combination of functions
$\Trr\mu^k=h_0^{k-1}(\gamma,\mu)$ and $\Trr\left(\mu^{k_1}P_+\mu^{k_2}P_+\ldots P_+\mu^{k_n}P_+\right)$. However due to the fact that $\dim\H_+=1$ we have
\be\Trr\left(\mu^{k_1}P_+\mu^{k_2}P_+\ldots P_+\mu^{k_n}P_+\right)=\frac{h_1^{k_1}(\gamma,\mu)}{k_1+1}\ldots \frac{h_1^{k_n}(\gamma,\mu)}{k_n+1}.\ee
Thus all integrals of motion $h_l^k$ are functionally dependent on $h_0^k$ and $h_1^k$. Therefore one has only two independent families of Hamilton equations, i.e. \eqref{h0} and \eqref{h1}.

%
In order to solve those families we note that
\be(\mu^k)_{-+}=M_k(\gamma,\mu)w ,\ee
\be(\mu^k)_{+-}=v^+M_k(\gamma,\mu),\ee
where
\be M_k(\gamma,\mu):=\!\!\left(\frac{h_1^{k-1}(\gamma,\mu)}{k}+\frac{h_1^{k-2}(\gamma,\mu)}{k-1}A+\ldots +
\frac{h_1^{2}(\gamma,\mu)}{3} A^{k-3}+a A^{k-2}+A^{k-1}\!\!\right)\ee
is a time independent operator.

Thus the equations \eqref{h0} assume the form
\be\left\{\begin{aligned}
&\frac\partial{\partial \tau_0^k}\, v=-\gamma\overline M_k(\overline\gamma,\mu^+)v\\
&\frac\partial{\partial \tau_0^k}\, w= \gamma M_k(\gamma,\mu)w\\
\end{aligned}\right..\ee
In this way we have reduced the system \eqref{h0} to linear system. Thus its solution is
\be v(\tau^2_0,\tau^3_0,\ldots)=\exp\left(\gamma\sum\limits_{k=2}^\infty M_k(\gamma,\mu^+(0,0,\ldots)) \tau^k_0\right) v(0,0,\ldots)\ee
\be w(\tau^2_0,\tau^3_0,\ldots)=\exp\left(-\gamma\sum\limits_{k=2}^\infty M_k(\gamma,\mu(0,0,\ldots)) \tau^k_0\right) w(0,0,\ldots)\ee
where $v(0,0,\ldots),w(0,0,\ldots)\in\ell^2$, $A\in L^\infty(\H_-)$, $a\in\C$ 
are initial conditions.

In the case of the equations \eqref{h1} we get
\be\left\{\begin{aligned}
&\frac\partial{\partial \tau_1^k}\, v=-\left(\sum_{j=1}^{k-1}\frac{\overline{h_1^{k-j}(\gamma,\mu)}}{k-j+1}M_{j-1}(\overline\gamma,\mu^+)+M_{k+1}(\overline\gamma,\mu^+)\right)v\\
&\frac\partial{\partial \tau_1^k}\, w=\left(\sum_{j=1}^{k-1}\frac{h_1^{j-1}(\gamma,\mu)}{j}M_{k-j}(\gamma,\mu)+M_{k+1}(\gamma,\mu)\right)w\\
\end{aligned}\right..\ee
These equation are also linear and their solution can be obtained by exponentiation.

\end{example}

\begin{example}[4-dimensional case]$\;$\label{ex:4}

In this example we solve the equation \eqref{ricatti} in the $\urtp$ case assuming $\dim\H_+=\dim\H_-=2$. We will use the following notation
\be \label{chi}\gamma=i\chi\qquad \mu=i\left(%
\begin{array}{cc}
  A & Z\\
  Z^+ & D\\
\end{array}%
\right),\ee
where $\chi\in\R$ and $A=A^+$, $D=D^+$, $Z\in Mat_{2\times2}(\C)$. Substituting \eqref{chi} into \eqref{ricatti} we obtain
\be \frac{d}{dt}A=0\qquad \frac{d}{dt}D=0\ee
and
\begin{align}\label{2x2}
 &\frac{d}{dt}Z=-i\chi(A^2Z+ZD^2+AZD+ZZ^+Z),\\
 &\frac{d}{dt}Z^+=i\chi(Z^+A^2+D^2Z^++DZ^+A+Z^+ZZ^+).
\end{align}
Let us note that equations \eqref{2x2} do not change their form with respect to the transformation $A\mapsto UAU^+$, $D\mapsto VD V^+$, $Z\mapsto VZ U^+$, where $UU^+=\1$ and $VV^+=\1$. So without the loss of the generality we can assume
\be A=\left(%
\begin{array}{cc}
  a_1 &0 \\
  0 & a_2\\
\end{array}\right),\quad
D=\left(%
\begin{array}{cc}
  d_1 & 0\\
  0 & d_2\\
\end{array}\right), \quad
Z=\left(%
\begin{array}{cc}
  a & b\\
  c & d\\
\end{array}\right),\ee
where $a_1,a_2,d_1,d_2\in\R$ are constants and $a,b,c,d$ are complex-valued functions of $t\in\R$. From \eqref{2x2} we obtain
\be\begin{split}\label{eq-abcd}
\frac{d}{dt}a&=i\chi\big(a_1^2+a_1d_1+d_1^2+\abs a^2+\abs b^2+\abs c^2\big)a+i\chi bc\bar d\\
\frac{d}{dt}b&=i\chi\big(a_1^2+a_1d_2+d_2^2+\abs a^2+\abs b^2+\abs d^2\big)b+i\chi a\bar cd\\
\frac{d}{dt}c&=i\chi\big(a_2^2+a_2d_1+d_1^2+\abs a^2+\abs c^2+\abs d^2\big)c+i\chi a \bar bd\\
\frac{d}{dt}c&=i\chi\big(a_2^2+a_2d_2+d_2^2+\abs b^2+\abs c^2+\abs d^2\big)c+i\chi \bar a  bc
   \end{split}\ee
Now we consider the generic case, i.e. $a_1\neq a_2$ and $d_1\neq d_2$.
In order to solve this system of equations we calculate explicitly the integrals of motion  $h_0^1(\mu)=\Trr \mu^2$, $h_0^2(\mu)=\Trr \mu^3$, $h_1^2(\mu)=\gamma\Tr(\mu^2P_+)$, $h_1^3(\mu)=\gamma\Tr(\mu^3P_+)$ and $h_0^3(\mu)=\Trr\mu^4$. From that we conclude that
\be \begin{split}
\abs a^2+\abs b^2&=:p^2=\const,\\
\abs a^2+\abs c^2&=:q^2=\const,\label{moduly}\\
\abs c^2+\abs d^2&=:r^2=\const,\\
\abs b^2+\abs d^2&=:s^2=\const,
    \end{split}\ee
\be \label{delta}\begin{split}
&\abs a^2 a_1d_1+\abs b^2 a_1d_2+\abs c^2 a_2d_1+\abs d^2 a_2d_2+\abs a^2\abs c^2+\abs b^2\abs d^2+2\Re(a\bar b \bar c d)\\
&=:\Delta=\const,\end{split}\ee
where $p^2+r^2=q^2+s^2$. Using \eqref{moduly} and \eqref{eq-abcd} we find
\be \label{eq:modul}\frac{d}{dt}\abs a^2=-\frac{d}{dt}\abs b^2=-\frac{d}{dt}\abs c^2=\frac{d}{dt}\abs d^2=2\chi \Im(a\bar b\bar c d)
\ee
Now from \eqref{delta} and \eqref{eq:modul} we obtain the following equation
\be
\frac{d}{dt}x =\pm\sqrt{w(x)}
\ee
on the function $x:=\abs a^2$, where
\be w(x):=4x(p^2-x)(q^2-x)(r^2-q^2-x)-v^2(x)\ee
and
\be\begin{split}
v(x):=&x(a_1-a_2)(d_1-d_2)-x(p^2-x)-(q^2-x)(r^2-q^2+x)-\\
-&a_1d_2p^2-a_2d_1q^2-a_2d_2(r^2-q^2).\end{split}\ee
Since $w$ is a polynomial of the fourth degree, this equation is solved by an elliptic integral of the first kind
\be t=\int \frac{dx}{\sqrt{w(x)}}.\ee
This allows us to express $x(t)$ as an elliptic function of time parameter $t$.

By \eqref{moduly} we may calculate $\abs b^2$, $\abs c^2$ and $\abs d^2$ in terms of $x(t)$.
In order to find the functions $a(t)$, $b(t)$, $c(t)$ and $d(t)$ we substitute their polar decompositions 
$a=\abs a e^{i\alpha}$, $b=\abs b e^{i\beta}$, $c=\abs c e^{i\gamma}$ and $d=\abs d e^{i\delta}$ into the equations \eqref{eq-abcd}. In this way we obtain
\be\begin{split}\label{fazy}
\frac{d}{dt}\alpha&=\chi\left(a_1^2+a_1d_1+d_1^2+p^2+q^2-x+\frac{v(x)}{2x}\right),\\
\frac{d}{dt}\beta&=\chi\left(a_1^2+a_1d_2+d_2^2+p^2+r^2-q^2+x+\frac{v(x)}{2(p^2-x)}\right),\\
\frac{d}{dt}\gamma&=\chi\left(a_2^2+a_2d_1+d_1^2+r^2+x+\frac{v(x)}{2(q^2-x)}\right),\\
\frac{d}{dt}\delta&=\chi\left(a_2^2+a_2d_2+d_2^2+p^2+r^2-x+\frac{v(x)}{2(r^2-q^2+x)}\right).
   \end{split}\ee
Since the right hand sides of the equations \eqref{fazy} are known, we can find $\alpha(t)$, $\beta(t)$, $\gamma(t)$ and $\delta(t)$ by integration. In this way we have solved the equations \eqref{2x2} in quadratures.

The Hamiltonian system solved in this example may have applications for example in the non-linear optics, see \cite{holm1}.
\end{example}

\appendix
\section{Extensions of Banach Lie groups and related Banach Lie--Poisson spaces}\label{ap-a}

Let us present an abbreviated description of extensions of Banach Lie groups, Banach Lie algebras and Banach Lie--Poisson spaces associated to them. Part of the results given below can be found in papers \cite{neeb-cext,neeb-ext-gr,michor-lie-alg,Oext}. Our main aim is to compute the formulas for the adjoint and coadjoint actions of extended Banach Lie group.

\subsection{Extensions of Banach Lie groups}$\;$

Let us consider an exact sequence of Banach Lie groups
\be\label{gr-ciag}\xymatrix@=35pt{
 {\set{e_N}}\ar@{^{(}->}[r]&  N  \ar@{^{(}->}[r]^-{\iota} & G \ar[r]^-{\pi} & H\ar[r] & {\set{e_H}}}.\ee
We assume that $N\to G\to H$ is a smooth principal bundle, i.e. the maps $\iota$ and $\pi$ are smooth and there exists a smooth local section $\sigma: U\to G$, where $U\subset H$ is an open neighborhood of identity.  Additionally we impose on $\sigma$ the normalization condition $\sigma(e_H)=e_G$. One can extend $\sigma$ to a global section
\be \label{section} \sigma: H\tto G,\ee
but in general such extension will not be smooth.
Let us define a map $\Psi: N\times H\tto G$ by
\be\Psi(n,h):=\iota(n)\sigma(h).\ee
Since $G/N\cong H$, we get that for $g\in G$ there exists a unique $n\in N$ such that $g=\iota(n) \sigma(\pi(g))$. Thus $\Psi$ is locally smooth bijection with inverse $\Psi^{-1}:G\tto N\times H$ given by 
\be \Psi^{-1}(g):=\big(\iota^{-1}(g\;(\sigma(\pi(g)))^{-1}),\sigma(\pi(g))\big).\ee

Using $\Psi$ one defines the multiplication on $N\times H$ by 
\be (n_1,h_1)\cdot (n_2,h_2) := \Psi^{-1}\big(\Psi(n_1,h_1)\Psi(n_2,h_2)\big)\ee
and can express it as follows
\be \label{gr-mult}(n_1,h_1)\cdot (n_2,h_2) =\big(n_1 \Phi(h_1)(n_2) \Omega(h_1,h_2),h_1h_2\big),\ee
where maps $\Phi:H \to \Aut(N)$ and $\Omega:H\times H\to N$ are defined by
\be \label{Phi}\Phi(h)(n):=\iota^{-1}\big(\sigma(h)\iota(n)\sigma(h)^{-1}\big),\ee
\be \label{Omega}\Omega(h_1,h_2):=\iota^{-1}\big(\sigma(h_1)\sigma(h_2)\sigma(h_1h_2)^{-1}\big).\ee
Let us denote by $\bar\Phi$ the map
\be\label{Phi-bar}\bar\Phi:H\times N\ni(h,n)\mapsto \Phi(h)(n)\in N.\ee

One has the following properties of $\Phi$ and $\Omega$:
\bse\label{ext-gr-cond}
\be \Phi(e_H)=\id\ee
\be \Omega(e_H,h)=\Omega(h,e_H)=e_N\ee
\be \Omega(h_1,h_2)\Omega(h_1h_2,h_3)=\Phi(h_1)\big(\Omega(h_2,h_3)\big)\Omega(h_1,h_2h_3)\ee
\be \Omega(h_1,h_2) \Phi(h_1h_2)(n)=\Phi(h_1)\circ\Phi(h_2)(n)\Omega(h_1,h_2)\ee\ese

Forgetting about the definitions \eqref{Phi}, \eqref{Omega} we can consider $\Phi$ and $\Omega$  as abstract maps satisfying the conditions \eqref{ext-gr-cond}. We assume that the map $\bar\Phi$ is smooth on $U\times N$ and $\Omega$ is smooth on some neighborhood of $(e_H,e_H)$. Moreover we have to assume that the map $H\ni x\mapsto \Omega(h,x)\Omega(hxh^{-1},h)^{-1}$ is smooth for all $h\in H$ on a neighborhood of $e_H$
(if $H$ is connected then this condition is automatically satisfied).
Under these conditions there exists on $N\times H$ a structure of Banach Lie group 
defined by \eqref{gr-mult}, see \cite{neeb-ext-gr}. One denotes this Banach Lie group by $N\times_{\Phi,\Omega}H$.

We get that the inverse of $(n,h)$ in $N\times_{\Phi,\Omega}H$ is given by
\be (n,h)^{-1}=(\Omega(h^{-1},h)^{-1}\Phi(h^{-1})(n^{-1}),h^{-1})\ee
and the inner automorphism $I_{(n,h)}(m,g):=(n,h)\cdot(m,g)\cdot(n,h)^{-1}$ can be expressed in terms of $\Phi$ and $\Omega$ as
\be \label{inner} I_{(n,h)}(m,g)=\big(n\Phi(h)(m) \Omega(h,g) \Omega(hgh^{-1},h)^{-1}\Phi(hgh^{-1})(n^{-1}),hgh^{-1}\big).\ee

Now, let us pass to the extensions of related Banach Lie algebras.

\subsection{Extensions of Banach Lie algebras}$\;$

We will denote the Banach Lie algebras of $G$, $H$, $N$ by $\g$, $\h$, $\n$ respectively. Taking derivatives
of the maps in \eqref{gr-ciag} we obtain the exact sequence of Banach Lie algebras
\be\label{alg-ciag}\xymatrix@=40pt{
 {\set{0}}\ar@{^{(}->}[r]&  \n  \ar@{^{(}->}[r]^-{D\iota(e_N)} & \g \ar[r]^-{D\pi(e_G)} & \h\ar[r] & {\set{0}}}.\ee

The derivative $D\Psi^{-1}(e_G):\g\to\n\oplus\h$ of $\Psi^{-1}$ at the point $e_G$ allows us to identify Banach space $\g$ with $\n\oplus\h$.

The adjoint representation of $N\times_{\Phi,\Omega} H$ on $\g\cong\n\oplus\h$ can be locally computed from \eqref{inner} and it is the following
\be\begin{split}\label{Ad-ext}
\Ad_{(n,h)}(\zeta,\eta)&=\bigg(\Ad_n\big(D_2\bar\Phi(h,e_N)(\zeta)+ D_2\Omega(h,e_H)(\eta) -  D_1\Omega(e_H,h)(\Ad_h \eta)\big) + \\
&+(DL_n(n^{-1})\circ D_1\bar\Phi(e_H,n^{-1})\circ\Ad_h)(\eta),\Ad_h \eta\bigg)
\end{split}\ee
for $(n,h)\in N\times_{\Phi,\Omega}U$, $(\eta,\zeta)\in \n\oplus\h$, where $\bar\Phi$ is defined by \eqref{Phi-bar},
and $D_i$ denotes partial derivative with respect to $i^{th}$ argument. We denote by $L_n$ the left group action $L_n m=nm$, $n,m\in N$, on itself. 

Differentiating \eqref{Ad-ext} we obtain the formula for Lie bracket 
\be\label{ext-bracket} [(\zeta,\eta),(\nu,\xi)]:=\big([\zeta,\nu]+\phi(\eta)(\nu)-\phi(\xi)(\zeta)+\omega(\eta,\xi),[\eta,\xi]\big),\ee
for $(\zeta,\eta),(\nu,\xi)\in \n\oplus\h$,
where $\phi:\h\to\Aut(\n)$ is the linear continuous map and $\omega:\h\times\h\to\n$ is the continuous bilinear skew symmetric map defined by $\Phi$ and $\Omega$ as follows:
\be \label{phi_d}\phi(\eta)(\zeta):=D_1D_2\bar\Phi(e_H,e_N)(\eta,\zeta),\ee
\be \label{omega_d}\omega(\eta,\xi):=D_1D_2\Omega(e_H,e_H)(\eta,\xi)-D_1D_2\Omega(e_H,e_H)(\xi,\eta).\ee
In these formulas $D_1D_2$ is second mixed partial derivative.



The maps $\phi$ and $\omega$ satisfy the following infinitesimal version of conditions \eqref{ext-gr-cond}:
\be\begin{split}\label{omega}
&\omega([\eta,\eta'],\eta'')+\omega([\eta',\eta''],\eta')+\omega([\eta'',\eta],\eta')-\\
&-\phi(\eta)(\omega(\eta',\eta''))-
\phi(\eta')(\omega(\eta'',\eta))-\phi(\eta'')(\omega(\eta,\eta'))=0,\end{split}\ee
and
\be \label{phi}\ad_{\omega(\eta,\eta')}+\phi([\eta,\eta'])-[\phi(\eta),\phi(\eta')]=0\ee
for all $\eta,\eta',\eta''\in\mathfrak h$.

If we forget about the underlying Banach Lie groups and consider their Banach Lie algebras only, then the maps $\phi$ and $\omega$ satisfying the conditions \eqref{omega}-\eqref{phi}
with additional smoothness conditions, define the structure of Banach Lie algebra on $\n\oplus \h$, see \cite{michor-lie-alg,Oext}.

\subsection{Extensions of Banach Lie--Poisson spaces}$\;$

According to \cite{OR} the \emphh{Banach Lie--Poisson space} is Banach space $\b$ such that its dual $\b^*$ is Banach Lie algebra with the property
\be \label{blp}\ad_x^* \b\subset \b\subset \b^{**}\ee
for all $x\in\b^*$. This property allows to define the Poisson bracket on $\b$
\be \label{pb-blp}\{ f,g\}(b)=\langle [Df(b),Dg(b)],b \rangle,\ee
where $Df(b),Dg(b)\in\b^*$ are Fr\'echet derivatives at point $b\in\b$. The bracket makes the Banach space $\b$ a Banach Poisson space in the sense of \cite{OR}.

Let us assume that Banach Lie algebras $\n$, $\h$ and $\g$ possess predual Banach spaces $\n_*$, $\h_*$ and $\g_*$ satisfying condition \eqref{blp}. We also assume that maps $(D\iota(e_N))^*$, $(D\pi(e_G))^*(h_*)$ dual to $D\iota(e_N)$ and $D\pi(e_G)$ preserve predual spaces, i.e.
\be (D\iota(e_N))^*(\g_*)\subset \n_*, \qquad (D\pi(e_G))^*(\h_*)\subset \g_*.\ee
In that situation one obtains the exact sequence of predual Banach spaces
\be\label{blp-ciag}\xymatrix@=40pt{
 {\set{0}}\ar@{^{(}->}[r]&  \h_*  \ar@{^{(}->}[r]^-{(D\pi(e_G))^*} & \g_* \ar[r]^-{(D\iota(e_G))^*} & \n_*\ar[r] & {\set{0}}},\ee
see Lemma 3.7 in \cite{Oext}.

We can identify $\g_*$ with $\n_*\oplus\h_*$ by the map dual to the derivative $D\Psi^{-1}(e_G)$ at the point $e_G$.
This identification allows us to compute coadjoint actions of  $N\times_{\Phi,\Omega}H$ and $\n\oplus\h$
on $\n_*\oplus\h_*$ as follows:
\be
\begin{split}
&\Ad^*_{(n,h)}(\tau,\mu)=\bigg((D_2\bar\Phi(h,e_N))^* \Ad_n^*\tau,\big((D_2\Omega(h,e_H))^*\Ad^*_n-\\
&-\Ad^*_h(D_1\Omega(e_H,h))^*\Ad^*_n+\Ad^*_h (D_1\bar\Phi(e_H,n^{-1}))^*\big)\tau+\Ad^*_h\mu\bigg),\end{split}\ee
where $(n,h)\in N\times_{\Phi,\Omega}U$, $(\tau,\mu)\in \n_*\oplus\h_*$ and
\be \label{coad-ext}\ad^*_{(\zeta,\eta)}(\tau,\mu)=\bigg(\ad^*_\zeta\tau+(\phi(\eta))^*\tau,
-(\phi(\,\cdot\,)(\zeta))^*\tau+(\omega(\eta,\,\cdot\,))^*\tau+\ad^*_\eta\mu\bigg)\ee
for $(\zeta,\eta)\in \n\oplus\h$, $(\tau,\mu)\in \n_*\oplus\h_*$.

The coadjoint representation \eqref{coad-ext} satisfies the condition \eqref{blp} if and only if
\be \label{coad-pres}(\phi(\eta))^*(\n_*)\subset\n_*\qquad (\phi(\,\cdot\,)(\zeta))^*(\n_*)\subset \h_*\qquad (\omega(\eta,\,\cdot\,))^*(\n_*)\subset\h_*.\ee
So, under these conditions the Banach space $\n\oplus\h$ is Banach Lie--Poisson space.
Using the definition \eqref{pb-blp} and Lie bracket\eqref{ext-bracket} we obtain the Poisson bracket on $\n_*\oplus\h_*$:
\be \begin{split}
\label{ext-pb}\{ f,g\}(\tau,\mu)&=\langle [D_1 f,D_1 g]+\phi(D_2 f)(D_1 g)-\phi(D_2 g)(D_1 f)+\omega(D_2 f,D_2 g) ,\tau \rangle+\\
&+\langle [D_2 f,D_2 g],\mu\rangle
    \end{split}\ee
for $f,g\in C^\infty(\n_*\oplus\h_*)$.

Further investigation of extensions of Lie groups and Lie algebras is beyond the scope of this paper, so for more information we refer to \cite{neeb-cext,neeb-ext-gr,michor-lie-alg,Oext}.

\section{Magri method}\label{ap:magri}

We briefly recall the Magri method of constructing of integrals of motion in involution. For more details see e.g. \cite{magri}.

One considers a pencil of compatible Poisson brackets
\be \pb_\epsilon:=\pb_1+\epsilon\pb_2 \ee
where $\epsilon\in\R$. 
Compatibility of Poisson brackets means that $\pb_\epsilon$ is also a Poisson bracket for any parameter $\epsilon$.
Let $I^k_\epsilon$ be a family of Casimirs for Poisson bracket $\pb_\epsilon$  indexed by $k\in\N$, i.e.
\be \label{casimir} \{I^k_\epsilon,\,\cdot\,\}_\epsilon=0.\ee

Assuming that $I^k_\epsilon$ depends analytically on the parameter $\epsilon$ one expands
the equality \eqref{casimir} and computes the  coefficients in front of $\epsilon^n$. Thus one obtains that $\{h_0^k,\,\cdot\,\}_1=0$ and
\be  \label{magri}\{h_{l}^k,\,\cdot\,\}_1=\{h_{l+1}^k,\,\cdot\,\}_2\qquad l=0,1,\ldots\ee
where $h_l^k$ are defined by
\be I^k_\epsilon=\sum_{l=0}^{\infty} h^k_l \epsilon^l. \ee
Due to the relation \eqref{magri}, sequence $\{h_l^k\}_{l\in\No}$ is called Magri chain.

By using \eqref{magri} twice one gets that
\be\{h_l^k,h_n^{k'}\}_1=\{h_{l-1}^k,h_{n+1}^{k'}\}_1.\ee
Next by iterating that procedure one concludes that
\be\{h_l^k,h_n^{k'}\}_1=\{h_{0}^k,h_{n+l}^{k'}\}_1=0\ee
Thus functions $h_l^k$ are in involution
\be \label{magri_inv}\{h_l^k,h_n^{k'}\}_\epsilon=\{h_l^k,h_n^{k'}\}_1=\{h_l^k,h_n^{k'}\}_2=0\ee
for all Poisson brackets under consideration.

\section*{Acknowledgments}
Authors would like to thank A.~B.~Tumpach, D.~Belti\c t\v a and V.~Dragovic for many valuable remarks and comments. This work is partially supported by Polish Government grant 1 P03A 001 29. Authors would like to thank Banach Center for its hospitality during the workshop "Banach Lie--Poisson spaces and integrable systems" (5-10 August 2008, B\c edlewo).

\end{document}